\documentclass[review]{elsarticle}

\usepackage{float}

\usepackage{lineno,hyperref}
\modulolinenumbers[5]

\journal{Journal of \LaTeX\ Templates}

\usepackage{graphicx}

\usepackage[cmex10]{amsmath}
\usepackage{amssymb}

\newcommand{\hl}[1]{{#1}} 

\usepackage[latin9]{inputenc}
\usepackage{float}
\usepackage{amsmath}
\usepackage{amssymb}
\usepackage{graphicx}
\usepackage{esint}

\makeatletter

\pdfpageheight\paperheight
\pdfpagewidth\paperwidth

\floatstyle{ruled}
\newfloat{algorithm}{tbp}{loa}
\providecommand{\algorithmname}{Algorithm}
\floatname{algorithm}{\protect\algorithmname}

\usepackage{color}

\usepackage{url}

\usepackage{algorithm}
\usepackage[noend]{algorithmic}

\makeatother

\begin{document}

\begin{frontmatter}

\title{Cascades in interdependent flow networks}
\author[cnr,imt,lims]{Antonio Scala\corref{mycorrespondingauthor}}
\cortext[mycorrespondingauthor]{Corresponding author}
\ead{antonio.scala@cnr.it}
\ead[url]{https://sites.google.com/site/antonioscalaphys/}

\author[gubkin]{Pier Giorgio De Sanctis Lucentini}

\author[imt,lims]{Guido Caldarelli}
\author[enea]{Gregorio D'Agostino}

\address[cnr]{ISC-CNR Physics Dept., Univ. dir Roma "La Sapienza", 00185 Roma, Italy}
\address[imt]{IMT Alti Studi Lucca, piazza S. Ponziano 6, 55100 Lucca, Italy}
\address[lims]{LIMS London Institute of Mathematical Sciences, 22 South Audley St Mayfair London UK}
\address[gubkin]{Gubkin Russian State University of Oil and Gas, Moskow, Russia}
\address[enea]{ENEA: Italian National Agency for New Technologies, Energy and Sustainable Economic Development, Rome, Italy}

\begin{abstract}
In this manuscript, we investigate the abrupt breakdown behavior of coupled distribution grids under load growth. \hl{This scenario mimics} the ever-increasing customer demand and the foreseen introduction of energy hubs interconnecting the different energy vectors.
We extend an analytical model of cascading behavior due to line overloads
to the case of interdependent networks and find evidence of first
order transitions due to the long-range nature of the flows. 
Our results indicate that the foreseen increase in the couplings between the grids has two competing effects: on the one hand, it increases the safety region where grids can operate without withstanding systemic failures; on the other hand, \hl{it increases the possibility of a joint systems' failure}.
\end{abstract}

\begin{keyword}
complex networks, interdependencies, mean field models
\end{keyword}

\end{frontmatter}


\section{Introduction}

Physical Networked Infrastructures ($PNI$s) such as power, gas or water distribution are at the heart \hl{of the functioning of our society}; they are very well engineered systems designed to be at least $N-1$ robust -- i.e., they should be resilient to the loss of a single component via automatic or human guided interventions. The constantly growing size of $PNI$s has increased the possibility of multiple failures which escape the $N-1$ criteria; however, implementing robustness to \textit{any sequence} of $k$ failures ($N-k$ robustness) requires an exponentially growing effort in means and investments. In general, since $PNI$s can be considered to be aggregations of a large number of simple units, they are expected to exhibit emergent behaviour, i.e. they show as a whole additional complexity beyond what is dictated by the simple sum of its parts \cite{AndersonScience1972}.

A general problem of $PNI$s are cascading failures, i.e. events characterized by the propagation and amplification of a small number of initial failures that, due to non-linearity of the system, assume system-wide extent. This is true even for systems described by linear equations, since most failures (like breaking a pipe or tripping a line) correspond to discontinuous variations of the system parameters, i.e. are a strong non-linear event. This is a typical example of emergent behavior leading
to one of the most important challenges in a network-centric word, i.e. systemic risk. An example of systemic risk in $PNI$s are the occurrence of blackout in one of the most developed and sophisticated
system, i.e. power networks. It is important to notice that if such large outages \hl{were} intrinsically due to an emergent behaviour of the electric power systems, increasing the accuracy of power systems' simulation \hl{would} not necessarily lead to better predictions of black-outs.

\hl{Power grids can be considered an example of complex networks}\cite{PaganiPhysA2013}\hl{and hence cascading failures in complex networks }\cite{Crucitti2004}\hl{ is field with important overlaps with system engineering and critical infrastructures protection; however, most of the cascading model are based on local rules that are not appropriate to describe systems like power grids }\cite{HinesCHAOS2010}\hl{ that, due to long range interactions, require a different approach }\cite{PahwaSciRep2014,ScalaIJCIS2015}\hl{.

Another import issue is increasing interdependent among critical infrastructures}\cite{Rinaldi2001}\hl{; seminal papers have pointed out the possibility of the occurrence of catastrophic cascades across interdependent networks }\cite{BuldyrevNAT2010,Brummitt2012}\hl{. However, there is still room for increasing the realism of such models}\cite{BookNON2014}\hl{, especially in the case of electric grids or gas pipelines. In this paper we move a preliminary step in such direction, trying to capture the systemic effect for coupled networks with long range interactions.}

To highlight the possibility of emergent behavior, we will first abstract $PNI$s in order to understand the basic mechanisms that could drive systemic failures; in particular, we will consider finite capacity networks where a commodity (a scalar quantity) is produced at source
nodes, consumed at load nodes and distributed as a Kirchoff flow (e.g. fluxes are conserved).
\hl{For such systems, we will first} introduce a simplified model that is amenable
of a self-consistent analytical solution. Subsequently, we will extended
such model to the case of several coupled networks and study the cascading
behavior under increasing stress (i.e. increasing flow magnitudes).

In section \ref{sec:model}, we develop our simplified model of overload
cascades first in isolated (sec. \ref{sec:1sys}) and coupled systems
(sec. \ref{sec:coupledsys}). In particular, in subsection \ref{sec:kirchoff},
we introduce the concept of flow network with a finite capacity and
relate conservation laws to Kirchoff's equations and to the presence
of long range correlation. \hl{To account for} such correlations, in subsection
\ref{sec:1sys} we introduce a mean field model for the cascade failures
of flow networks; in subsection \ref{sec:coupledsys}, we extend the
model to the case of several interacting systems. Finally, in section
\ref{sec:discuss} we discuss and summarize our results.

\section{Model}

\label{sec:model}

\subsection{Flow networks}

\label{sec:kirchoff}

Let's consider a network $G=\left(\mathcal{V},\mathcal{E},\mathbf{c}\right)$
where $\mathcal{V}=\left\{ 1\leq i\leq\left|\mathcal{V}\right|\right\} $
is the node set, $\mathcal{E}\subseteq\mathcal{V}\times\mathcal{V}$
is the set of edges and $\mathbf{c}=\{c_{\left(i,,j\right)}\}$ is
the vector characterizing the capacities of the edges $\left(i,j\right)$.
We associate \hl{to} the nodes a vector $\mathbf{p}=\{p_{i}\}$ that characterize
the production ($p_{i}>0$) or the consumption ($p_{i}<0$) of a commodity.
We further assume that there are no losses in the network (i.e. $\sum_{i}p_{i}=0$);
hence, the total load on the network is

\[
L={\displaystyle \sum_{i:p_{i}>0}}p_{i}
\]

The distribution of the commodity is described by the fluxes $\mathbf{f}=\{f_{\left(i,j\right)}\}$
on the edges $\left(i,j\right)\in\mathcal{E}$ \hl{that are} supposed to
respect Kirchoff equations, i.e.

\begin{equation}
\sum_{j}f_{\left(i,j\right)}=p_{i}\label{eq:Kirchoff}
\end{equation}

The relation among fluxes and demand/load is described by constitutive
equations 
\begin{equation}
\mathbf{f}=F(\mathbf{p},G)\label{eq:constitutive}
\end{equation}
where in general eq.\ref{eq:constitutive} is non-linear but satisfies
eq.\ref{eq:Kirchoff}.

The finite capacity $c_{\left(i,j\right)}$ constrains the maximum
flux on link $\left(i,j\right)$ 
\begin{equation}
|f_{\left(i,j\right)}|<c_{\left(i,j\right)}
\end{equation}
above which the link will cease functioning. As an example, power
lines are tripped (disconnected) when power flow goes beyond a certain
threshold. Since flows will redistribute after a link failure, it
could happen that other lines get above their flow threshold and hence
consequently fail, eventually leading to a cascade of failures. A
typical algorithm to calculate the consequences of an initial set
of line failures $\mathcal{F}^{0}=\{(ij)\,\,$failed$\}$ is the alg.\ref{alg:net_cascade}.

\begin{algorithm}
\caption{Network cascading}

\begin{algorithmic} 

\STATE Set initial failures $\mathcal{F}^{0}$ 

\STATE $t\leftarrow0$ 

\REPEAT 

\STATE $t\leftarrow t+1$ 

\STATE Calculate flows $\mathbf{f}^{t}\leftarrow F(\mathbf{p},G|\mathcal{F}^{t-1})$ 

\STATE Calculate new failures $\Delta\mathcal{F}^{t}\leftarrow\{(ij):|f_{ij}^{t}|>c_{ij}\}$ 

\STATE $\mathcal{F}^{t}\leftarrow\mathcal{F}^{t-1}\cup\Delta\mathcal{F}^{t}$ 

\UNTIL$\Delta\mathcal{F}^{t}\equiv\emptyset$ 

\end{algorithmic} 

\label{alg:net_cascade}
\end{algorithm}

Here $F(\mathbf{p},G|\mathcal{F})$ calculates the flows subject to
the constrains that flows are zero in the failure set of edges $(i,j)\in\mathcal{F}$.

To develop a general model that helps us understanding the class of
failures that can affect Kirchoff-like flow networks, let's start
from rewriting eq.\ref{eq:Kirchoff} in matrix form 

\begin{equation}
\mathcal{B^{T}}\mathbf{f}=\mathbf{p}\label{eq:MatrixKirchoff}
\end{equation}
using the incidence matrix $B$ that associates to each link $\left(i,j\right)$
its nodes $i$ and $j$ and vice-versa. $B$ is an $\left|\mathcal{V}\right|\times\left|\mathcal{E}\right|$
matrix where each column corresponds to an edge $\left(i,j\right)$;
its columns are zero-sum and the only two non-zero elements have modulus
$1$ and are on the $i^{th}$ and on the $j^{th}$ row.

The matrix $B$ is related to the Laplacian $B^{T}B$ of the system;
in particular, it shares the same right eigenvalues and the same spectrum
(up to a squaring operations); hence, it is a long-range operator
since perturbation on a node of the system can be reflected on nodes
far away on the network \cite{PahwaSciRep2014,ScalaIJCIS2015}.

\subsection{Mean field model for cascades on a single network}

\label{sec:1sys}

Due to the long range nature of Kirchoff's equations, to understand
the qualitative behavior of such networks we can resort to a mean
field model of flow networks where one assumes that when a link fails,
its flow is re-distributed equally among all other links. 
Subsequently, the lines above their threshold would trip again, their flows would
be re-distributed and so on, up to convergence; recalling that $L$
is the total load of the system and assuming the each link $\left(i,j\right)$
has an initial flux $f=L/\left|\mathcal{E}\right|$, we can describe
such a model by alg.\ref{alg:MFcascading}.
\hl{Such model, introduced in }\cite{PahwaSciRep2014}\hl{, is akin to the fiber-bundle model }\cite{Peirce1926,Daniels1945}\hl{ and has been considered in more details in }\cite{Scala2015arXiv150601527S,YaganPRE2015}\hl{ for the case of a single system. While similar in spirit to the CASCADE model for black-outs }\cite{DobsonHAWAI2002,DobsonPEIS2005}\hl{, it yelds different results since it does not describe the statistic of the cascades in power systems but concentrates on the order of the transition in a single system.}

\begin{algorithm}
\caption{Mean Field cascading}

\begin{algorithmic}

\STATE $t\leftarrow0$ 

\STATE $F^{t}\leftarrow0$ initial number of failed links 

\REPEAT 

\STATE $t\leftarrow t+1$ 

\STATE $M\leftarrow\left|\mathcal{E}\right|-F^{t-1}$ number of working
links 

\STATE $l\leftarrow L/M$ average flux on the working links

\STATE $F^{t}\leftarrow\left|\left\{ (ij):l>c_{(ij)}\right\} \right|$ 

\UNTIL$F^{t}=F^{t-1}$

\end{algorithmic} 

\label{alg:MFcascading}
\end{algorithm}
Such algorithm can be cast in the form of a single equation in the
case where the system is composed by a large number of elements with
capacity $c$. In fact, in such limit we can describe the links' population
by the probability \hl{distribution} function $p(c)$ of their capacities. Indicating
with $M=\left|\mathcal{E}\right|$ the initial number of links, we
see that if we apply an overall load $L$ to the system, all the links
will be initially subject to a flow $l^{0}=L/M$. Thus, a fraction
of links $f^{1}=\int_{0}^{L/M}p\left(c\right)dc$ would immediately
fail, since their thresholds are lower than the flux $l$ they should
sustain. After the first stage of a cascade, there will be $M^{1}=(1-f^{1})M$
surviving links and the new load per link is $l^{1}=L/M^{1}$. The
following cascade's stages follow analogously; we can thus write the
mean field equations for the $\left(t+1\right)^{th}$ stage of the
cascade: 
\begin{equation}
f^{t+1}=P\left(\frac{l}{1-f^{t}}\right)\label{eq:MFmodel1sys}
\end{equation}
where $l=L/M$ is the initial load per link and $P\left(x\right)=\int_{0}^{x}p\left(c\right)dc$
is the cumulative distribution function of link capacities; the initial
conditions are $f^{t=0}=0$. The fix-point $f^{*}$ of eq.\ref{eq:MFmodel1sys}
satisfies the equation 

\begin{equation}
f^{*}=P\left(\frac{l}{1-f^{*}}\right)\label{eq:fixpoint1sys}
\end{equation}

and represents the total fraction of links broken at the end of the
cascading stages \cite{PahwaSciRep2014,ScalaIJCIS2015}.

The behavior of $f^{*}$ depends on the functional form of $p\left(c\right)$.
In particular, \hl{following}\cite{daSilveiraPRL1997}\hl{ we can define} $\pi(c)=1-P(c)$ and $x=l^{-1}(1-f)$ and we have that 
\begin{equation}
f=\int_{0}^{1/x}p\left(c\right)dc=1-\pi\left(\frac{1}{x}\right)
\end{equation}
so that we can rewrite eq.(\ref{eq:MFmodel1sys}) as 
\begin{equation}
lx^{t+1}=\pi\left(\frac{1}{x^{t}}\right)
\label{eq:dasilveira}
\end{equation}
\hl{Equation} (\ref{eq:dasilveira}) has a trivial fix-point $x^{*}=0$ (representing
a total breakdown of the system) since $\pi\left(\infty\right)=0$.
Such fix-point is unstable for $l\to0$ and becomes stable for $l>\partial_{x}\pi(x^{-1})|_{x\to0}$.
We notice that if $P\left(c\right)$ does not change convexity (i.e.
has no bumps) and the transition is first order, the system will breakdown
directly to the total collapsed state $f=1$.

In general, the behavior of the fix-point $x^{*}$ depends on the tail of the distribution $p(c)$ and is known to present a first order transition for a wide family of curves \cite{daSilveiraPRL1997}.

Depending on the functional form of $p\left(c\right)$, eq.(\ref{eq:fixpoint1sys}) could sometimes be solved analytically. Otherwise, the fix-point of eq. (\ref{eq:fixpoint1sys}) can be solved numerically either by iterating the eq. (\ref{eq:MFmodel1sys}) or by finding the zeros of eq.(\ref{eq:fixpoint1sys}) by Newton-Raphson iterations.

\hl{Notice that, if the system is long range, modelling cascade via homogeneous load redistribution allows to capture the order of the transition even when it gives not an accurate prediction of the actual location of the transition point. An example of such accordance for the case of power networks is given in }\cite{PahwaSciRep2014,ScalaIJCIS2015}\hl{, where both synthetic networks, realistic networks and mean-field systems show a first order transition.}

\subsection{Mean field model for interacting cascades}

\label{sec:coupledsys}

Commodities are defined substitutable when they can be used for the same aim; when commodities are substitutable, they can expressed in the same units. An example of such commodities are electricity and gas, \hl{since both can be used} for domestic heating. Hence, an increase on the cost of the gas (as the one that has been recently experienced by Ukraine) could induce stress on the electric network of the country since most customer will possibly switch to the cheaper energy vector \footnote{\hl{\textit{energy vectors} are man-made forms of energy that enable energy to be carried and can then be converted back into any other form of energy}}.
To take account for such effects, we will extend the model described
by eq.(\ref{eq:MFmodel1sys}) to the case of several coupled systems
that transport substitutable commodities.

We will consider $n$ coupled systems assuming that when a system $a$ is subject to some failures, it sheds a fraction $T_{a\to b}$ of the \hl{induced flow increase} on system $b$. In
other words, after failure system $a$ decreases its stress by a quantity $l_{a}f_{a}{\displaystyle \sum_{b\neq a}}T_{a\to b}$ by increasing the load of all other systems $b\neq a$ by $l_{a}f_{a}T_{a\to b}$. Thus, the $n$ coupled systems are described by a set of $n$ equations of the form of eq.(\ref{eq:MFmodel1sys}) 
\begin{equation}
f_{a}^{t+1}=P_{a}\left(\frac{\tilde{l_{a}^{t}}}{1-f_{a}^{t}}\right)\label{eq:MFmanysys}
\end{equation}
where $\tilde{l_{a}^{t}}$ is the load per link experimented by system
$a$ in the $t^{th}$ stage of the cascade and $P_{a}\left(x\right)=\int_{0}^{x}p_{a}\left(x\right)dx$
is the cumulative of the probability distribution function $p_{a}\left(x\right)$
for the capacities of the $a^{th}$ system. Equations (\ref{eq:MFmanysys})
are not independent, since the systems' coupling is reflected by the
dependence of $\tilde{l_{a}^{t}}$ on the fractions $f_{b}^{t}$ of
failed links in all the other systems, i.e. 
\begin{equation}
\tilde{l_{a}^{t}}=l_{a}(1-f_{a}^{t}\sum_{b}T_{a\to b})+\sum_{b}T_{b\to a}l_{b}f_{b}^{t}=l_{a}+\sum_{b}\mathcal{L}_{ab}l_{b}f_{b}^{t}\label{eq:MF2coupl}
\end{equation}
where $\mathcal{L}_{ab}=(1-\delta_{ab})T_{b\to a}+\delta_{ab}\sum_{b}T_{a\to b}$
has the form of a Laplacian operator. Thus, the full equations for
$n$ coupled systems are 
\begin{equation}
f_{a}^{t+1}=P_{a}\left(\frac{l_{a}+\sum_{b}\mathcal{L}_{ab}l_{b}f_{b}^{t}}{1-f_{a}^{t}}\right)\label{eq:MFmanysysFull}
\end{equation}
. 

\begin{figure}[H]
\begin{centering}
\includegraphics[width=1\columnwidth]{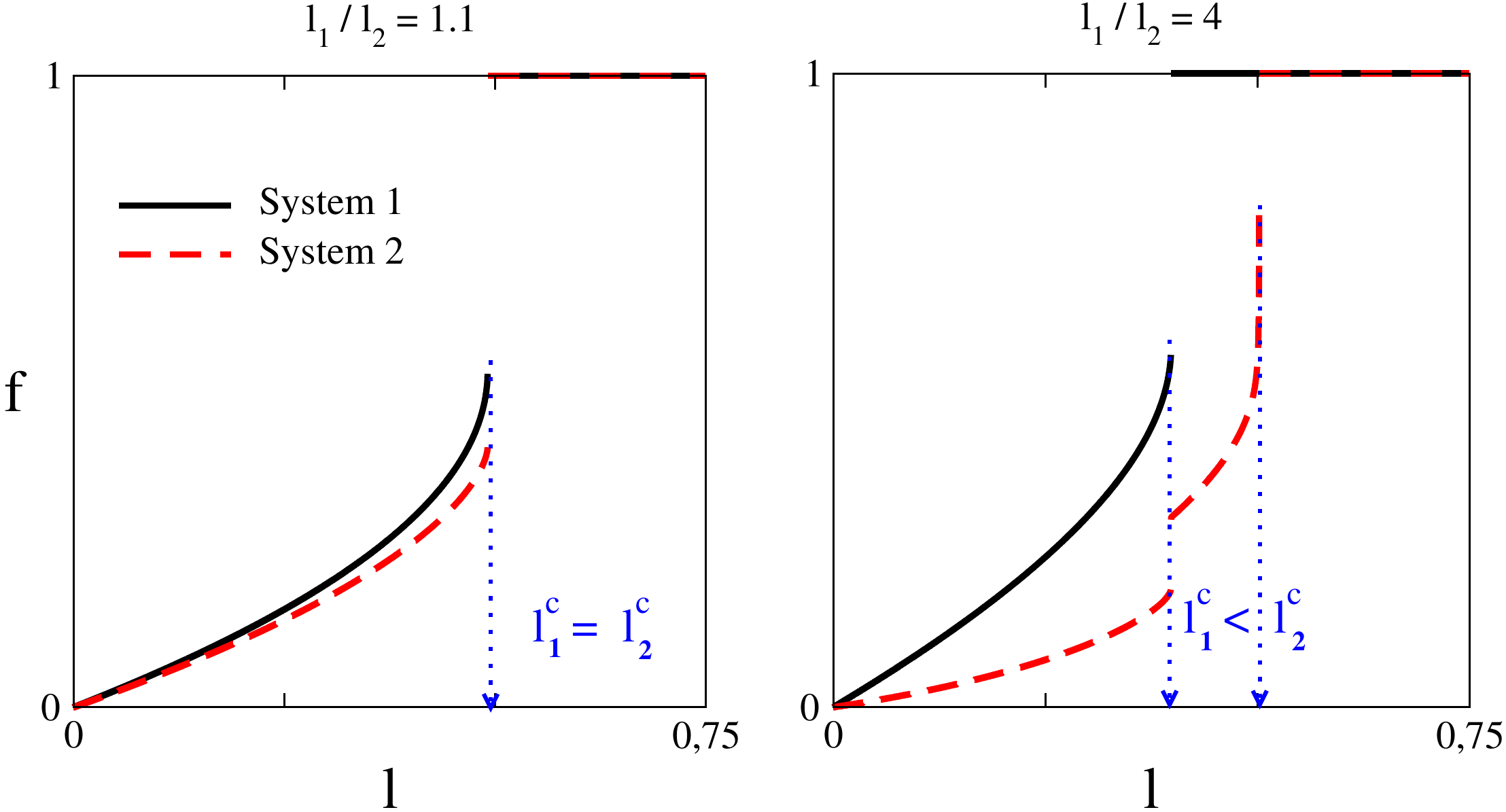} 
\par\end{centering}

\caption{Behaviour of the number of failed nodes respect to the total stress
$l=l_{1}+l_{2}$ of the systems. For simplicity, we present the case
of two identical systems with a flat distribution of link capacities
and symmetric couplings $T_{1\to2}=T_{2\to1}=0.5$. We show the result
of increasing the total stress $l$ in the two systems along the lines
$l_{1}/l_{2}=const$. \textbf{Left panel}: we show the case $l_{1}/l_{2}=1.1$
where both systems are subject to a similar stress while increasing
$l$. In such case both system break down together at the same critical
load $l_{1}^{c}=l_{2}^{c}$; in the region $l>l_{1}^{c}=l_{2}^{c}$
both systems \hl{have} failed. \textbf{Right panel}: we show the case $l_{1}/l_{2}=4$
where when increasing $l$ system $1$ is more stressed than system
$2$. In this case, the break down of system $1$ at the critical
load $l_{1}^{c}$ induces a jump in the number of failures system
$2$, but system $2$ is still able to sustain stress and will break
down only at higher values of $l$. Respect to the $l_{1}\sim l_{2}$
case, there is now a region $l_{1}^{c}<l<l_{2}^{c}$ where only system
$1$ \hl{has} failed. }

\newpage

\label{fig:2behaviours} 
\end{figure}

For simplicity, \hl{from now on} we will consider the case of two identical system\hl{s}
with a uniform distribution of link capacities. \hl{Notice that for a single system the transition is first order unless the probability distribution of the capacities is a power-law }\cite{daSilveiraPRL1997} \hl{-- an event that is not realistic for real world flow networks.
Since the functional form of $P(.)$ is easy to recover for a uniform distribution, we can solve the fix-point of eq.(}\ref{eq:MFmanysysFull}\hl{) numerically by iterating the equations up to convergence; an alternative methodology would be using Newton-Raphson algorithms.}
We show in fig.(\ref{fig:2behaviours})
the cascading behavior of two coupled systems; we observe that --
as in the single system case -- transitions are in the form of abrupt
jumps, i.e. are first order. Let's rewrite eq.(\ref{eq:MFmanysysFull})
in the case of symmetric couplings $T_{1\to2}=T_{2\to1}=1$ and same
probability distribution for the capacities 
\begin{equation}
\left\{ \begin{array}{c}
f_{1}^{t+1}=P\left(\frac{l_{1}}{1-f_{1}^{t}}\left[1-T\left(f_{1}-\frac{l_{2}}{l_{1}}f_{2}\right)\right]\right)\\
f_{2}^{t+1}=P\left(\frac{l_{1}}{1-f_{2}^{t}}\left[1-T\left(f_{2}-\frac{l_{1}}{l_{2}}f_{1}\right)\right]\right)
\end{array}\right.\label{eq:MF2sysSymm}
\end{equation}

If the two systems described by eq.(\ref{eq:MF2sysSymm}) are stressed
at the same pace (i.e. $l_{1}=l_{2}=l/2$), we get the case 
\[
\left\{ \begin{array}{c}
f_{1}^{t+1}=P\left(\frac{l}{1-f_{1}^{t}}\left[1-T\Delta f_{12}\right]\right)\\
f_{2}^{t+1}=P\left(\frac{l}{1-f_{2}^{t}}\left[1+T\Delta f_{12}\right]\right)
\end{array}\right.
\]
; from the symmetric solution $\Delta f_{12}=0$ we see that the breakdown
of both systems happen at the same critical load as the uncoupled
systems. Such situation is shown in the left panel of fig.(\ref{fig:2behaviours}).

In the general, only one of the systems will be the first one to break
down (i.e. the fraction of broken links jumps to $f^*=1$): correspondingly,
also the other systems will experience a jump in the number of broken
links. Let's consider the symmetric case described by equations (\ref{eq:MF2sysSymm})
and suppose that $l_{1}>l_{2}$, so that system $1$ is the first
to breakdown (i.e. $f_{1}^*=1$); hence, the equation for the fix-point
of the second system becomes
\[
f_{2}^{*}=P\left(\frac{l}{1-f_{2}^{*}}\left[1+T\left(1-f_{2}^{*}\right)\right]\right)=P\left(\frac{l^{+}}{1-f_{2}^{*}}\right)
\]

i.e. the system behaves like a single system starting with a renormalized load $l^{+}=l\,\left[1+T\left(1-f_{2}^{*}\right)\right]>l$. Thus, if $l^{+}<l_{c}$ ( the critical value of eq.(\ref{eq:MFmodel1sys}), system $2$ will break down at higher values of the stress. Such situation is shown in the right panel of fig.(\ref{fig:2behaviours}).

\begin{figure}[H]
\begin{centering}
\includegraphics[width=1\columnwidth]{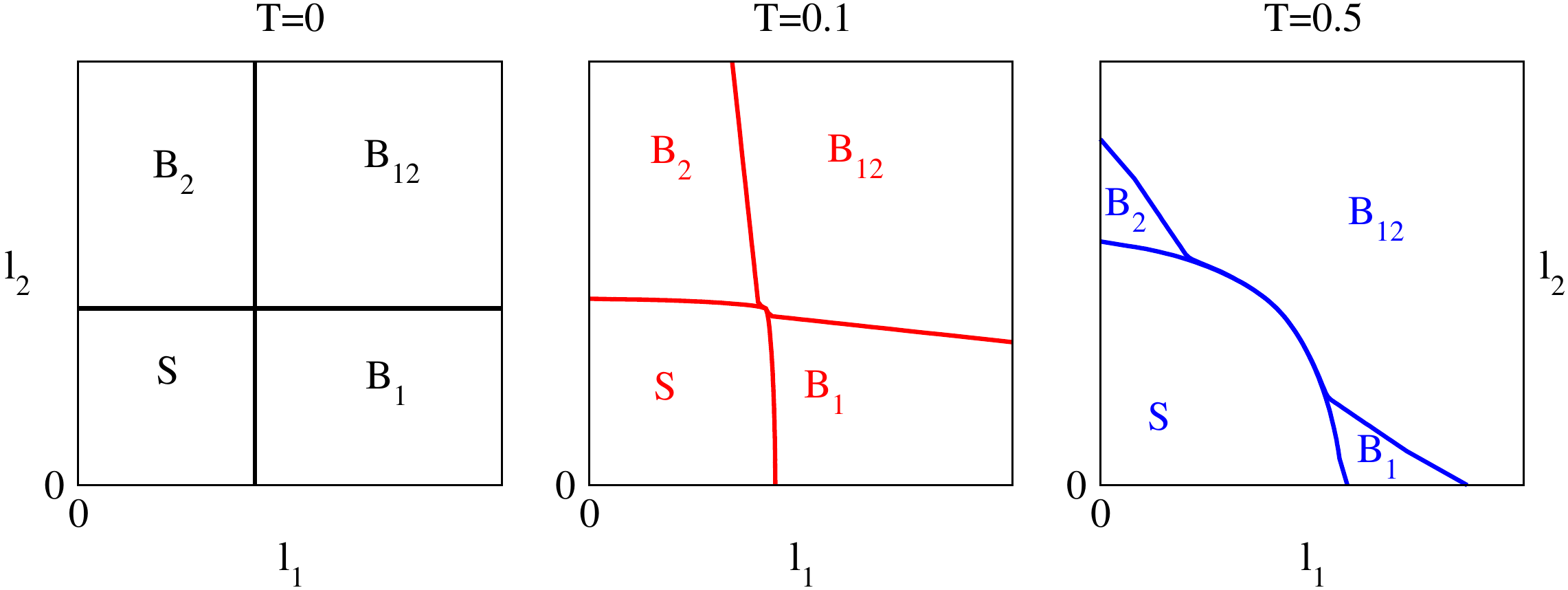} 
\par\end{centering}

\caption{Phase diagrams of two identical coupled systems with symmetric interactions
($T_{1\to2}=T_{2\to1}=T$). The plane of initial loads $l_{1}$ and
$l_{2}$ is separated in four different regions by critical transition
lines. The labels $B_{i}$ ($i=1,2$) mark the areas where only system
$i$ suffers systemic cascades ($f_{i}^*=1$,$f_{j\neq i}^*<1$), while
the label $B_{12}$ marks the area where both systems suffer system
wide cascades ($f_{1}^*=f_{2}^*=1$). The label $S$ marks the area near
the origin where no systemic cascades occur. \textbf{Left panel:}
the case $T=0$ corresponds to two uncoupled systems: thus, each system
suffers systemic failure at $l_{i}>l^{c}$ (where $l^{c}$ is the
critical load for an isolated system); both systems \hl{have} failed in
the $B_{12}$ area corresponding to the quadrant $(l_{1}>l^{c},l_{2}>l^{c})$.
\textbf{Central panel, right panel:} when couplings are introduced,
each system is able to discharge stress on the other one and the area
$S$ where both systems are safe increases. On the other hand, the
area $B_{12}$ where \emph{both} systems \hl{are in a failed state} increases. }

\label{fig:PhaseDiagrams} 
\end{figure}

In fig.(\ref{fig:PhaseDiagrams}) we show the full phase diagrams
of two coupled systems while varying the coupling among them. 
According to the initial loads, we can distinguish an area $S$ near the origin where the system is safe and three separate cascade regimes: $B_{1}$ and $B_{2}$, where either system $1$ or $2$ fails, and $B_{12}$ where both systems fail. 
We notice that, by increasing the coupling among the systems, both the area $S$ where the two systems are safe and the area $B_{12}$ where they fail together grow; accordingly, the areas $B_{i}$ where only one system fails shrink.

\section{Discussion}

\label{sec:discuss}

We have introduced a model for cascade failures due to the redistribution of flows upon overload of link capacities. For such a model, we have developed a mean field approximation both for the case of a single network and for the case of coupled networks. Our model is inspired
to a possible configuration for future power systems where network nodes the so-called energy hubs \cite{Geidl2007}, i.e. points where several energy vectors converge and where energy demand/supply can
be satisfied converting one kind of \hl{energy} in another. Hubs condition, transform and deliver energy in order to cover consumer needs \cite{Perrod2005}. In such configurations, one can alleviate the stress on a network by using the flows of the the other energy vectors; on the other hand, transferring loads from a network to the other can trigger cascades that can eventually backfire.

By analyzing the case of two coupled systems and by varying the strength
of the interactions among them, we have shown that at low stresses
coupling has a beneficial effect since some of the loads are shed
to the other systems, thus postponing the occurrence of cascading
failures. On the other hand, with the introduction of couplings the
region where not only one system fails but both systems fail together
also increases. The higher the couplings, the more the two systems
behave like a single one and the area where only a system \hl{has} failed
shrinks. 

\hl{Our model also applies to the realistic scenario where existent grids gets connected to allow power to be delivered across states; such scenario has inspired the analysis of }\cite{Brummitt2012}\hl{ that, even using an unrealistic model of power redistribution in electric grids, reaches conclusion that are similar to ours.}

It is worth noting that while fault propagation models do predict a general lowering of the threshold for coupled systems \cite{WangDAgostinoPRE2013}, in the present model a beneficial effect due to the existence of the interdependent networks is observed for small enough overloads, while the expected cascading effects take place only for large initial disturbances.
This picture is consistent with the observed phenomena for interdependent Electric Systems. Moreover the existence of interlinks among different networks may increase their synchronization capabilities \cite{MartinHernandez201492}.

\section*{Acknowledgements}
AS and GD acknowledge the support from EU HOME/2013/CIPS/AG/4000005013 project CI2C.
AS acknowledges the support from  CNR-PNR National Project "Crisis-Lab". 
AS and GC acknowledge the support from EU FET project DOLFINS nr 640772
and EU FET project MULTIPLEX nr.317532.
GD acknowledges the support from FP7 project n.261788 AFTER.

Any opinion, findings and conclusions or reccomendations expressed in this material are those of the author(s) and do not necessary reflect the views of the funding parties.

\section*{References}

\newpage

\end{document}